%% file: main.tex
\documentclass[sigconf]{acmart}
\makeatletter
\def\@ACM@checkaffil{% Only warnings
    \if@ACM@instpresent\else
    \ClassWarningNoLine{\@classname}{No institution present for an affiliation}%
    \fi
    \if@ACM@citypresent\else
    \ClassWarningNoLine{\@classname}{No city present for an affiliation}%
    \fi
    \if@ACM@countrypresent\else
        \ClassWarningNoLine{\@classname}{No country present for an affiliation}%
    \fi
}
\makeatother

\usepackage[ruled,linesnumbered,noend]{algorithm2e}
\usepackage{subfigure}

\usepackage{amsfonts,amssymb}
\usepackage{multirow}
\usepackage{enumitem} 
\usepackage{threeparttable}
\usepackage{balance}
\usepackage[normalem]{ulem}
\usepackage{amsmath,bm}
\usepackage{enumitem}
\usepackage{balance}
\usepackage{threeparttable}
\usepackage{color}
\usepackage{url}
\usepackage{appendix}

\newtheorem{myDef}{Definition}
\usepackage{graphicx}
\usepackage{flushend}
\usepackage{balance}
\usepackage{amsmath,amssymb,amsfonts,bm}
\usepackage{upgreek}
\usepackage{bbding}
\usepackage{pifont}
\usepackage{gensymb}
\usepackage{subfigure}
\usepackage{array}
\usepackage{colortbl}
\usepackage{booktabs}
\usepackage{pifont}
\usepackage{makecell}
\usepackage{amsthm}
\newcommand{\cmark}{\ding{51}} % ✓
\newcommand{\xmark}{\ding{55}} % ✗
\newcommand{\parhead}[1]{\noindent\textbf{#1}}

\newcommand{\down}{\vspace*{0.03in}}
\newcommand{\emp}[1]{%
\noindent\underline{\textit{#1}}}

\usepackage[table]{xcolor} 
\usepackage{collcell}      
\usepackage{pgf}     
\usepackage[most]{tcolorbox}
\usepackage{listings}
\definecolor{lowcolor}{HTML}{FF9999}  % [低分] 浅红
\definecolor{midcolor}{HTML}{FFFFFF}  % [中点] 纯白 (您希望60分是这个颜色)
\definecolor{highcolor}{HTML}{6FA8DC} % [高分] 矢车菊蓝 

\definecolor{lightgray}{gray}{0.9}

\newcommand{\Takeaway}[1]{%
     \noindent % 增加一点行间距并取消首行缩进
    \colorbox{lightgray}{\textbf{Takeaway #1:}}%
}
 
\usepackage[most]{tcolorbox}
\usepackage{xcolor}

\newtcolorbox{takeaway}{
    colback=gray!15, 
    colframe=white,
    before skip=0pt,       % 设为0，它将遵循正文的 \parskip
    after skip=0pt,        
    boxsep=0pt,            % 核心：消除全局内填充
    left=0pt, 
    right=0pt, 
    top=0pt,               % 设为0，底色高度将完全等于文字行高
    bottom=0pt,
    boxrule=0pt,
    sharp corners,         % 必须用直角，否则圆角会产生视觉空隙
    lines before=0,
    lines after=0,
    breakable,
    enhanced,
}
\newcommand{\ApplyGradient}[1]{%
  \if\relax\detokenize{#1}\relax%
  \else%
    \pgfmathsetmacro{\Raw}{#1}%
    \ifdim\Raw pt > 100pt \pgfmathsetmacro{\Raw}{100} \fi%
    \ifdim\Raw pt < 0pt \pgfmathsetmacro{\Raw}{0} \fi%
    
    \pgfmathsetmacro{\MidPoint}{60} 
    \ifdim\Raw pt < \MidPoint pt%
      \pgfmathsetmacro{\Ratio}{\Raw / \MidPoint * 100}%
      \edef\x{\noexpand\cellcolor{midcolor!\Ratio!lowcolor}}\x%
    \else%
      \pgfmathsetmacro{\Ratio}{(\Raw - \MidPoint) / (100 - \MidPoint) * 100}%
      \edef\x{\noexpand\cellcolor{highcolor!\Ratio!midcolor}}\x%
    \fi%
    #1%
  \fi%
}

\newcolumntype{H}{>{\collectcell\ApplyGradient}c<{\endcollectcell}}
\newtcolorbox{findingbox}{
  breakable,            
  colback=gray!15,        
  colframe=black,      
  boxrule=0.5pt,         
  arc=0mm,                
  left=1mm,              
  right=1mm,             
  top=1mm,              
  bottom=1mm,           
}

\makeatletter
\newcommand{\adaptiveTitle}[1]{%
  \ifdim\f@size pt > 12pt%
    \scalebox{0.96}[1]{#1}%
  \else
    \begingroup
      \tolerance=2000      % 提高容忍度
      \emergencystretch=20pt % 允许额外的拉伸空间
      #1%
    \endgroup
  \fi
}
\makeatother

\AtBeginDocument{%
  \providecommand\BibTeX{{%
    \normalfont B\kern-0.5em{\scshape i\kern-0.25em b}\kern-0.8em\TeX}}}

\copyrightyear{2026}
\acmYear{2026}
\setcopyright{cc}
\setcctype{by}
\acmConference[KDD 2026] {Proceedings of the 32nd ACM SIGKDD Conference on Knowledge Discovery and Data Mining V.2}{August 9--13, 2026}{Jeju Island, Republic of Korea.}
\acmBooktitle{Proceedings of the 32nd ACM SIGKDD Conference on Knowledge Discovery and Data Mining V.2 (KDD 2026), August 9--13, 2026, Jeju Island, Republic of Korea}
\acmISBN{979-8-4007-2259-2/2026/08}
\acmDOI{10.1145/3770855.3817516}
\begin{document}

%%
%% The "title" command has an optional parameter,
%% allowing the author to define a "short title" to be used in page headers.

\title{{DTBench: A Synthetic Benchmark for Document-to-Table Extraction}}

% \title{%
%   \texorpdfstring{%
%     \adaptiveTitle{DTBench: A Synthetic Benchmark for Document-to-Table Extraction}%
%   }{DTBench: A Synthetic Benchmark for Document-to-Table Extraction}%
% }
 
%% The "author" command and its associated commands are used to define
%% the authors and their affiliations.
%% Of note is the shared affiliation of the first two authors, and the
%% "authornote" and "authornotemark" commands
%% used to denote shared contribution to the research.

%%%%%% To Save Space
% \author{
% Yuxiang Guo$^{1}$, Zhuoran Du$^{1}$, Nan Tang$^{2}$, Kezheng Tang$^{1}$, Congcong Ge$^{1}$,  Yunjun Gao$^{1}$\\
% \normalsize $^{1}${Zhejiang University, Hangzhou, China} \\
% \normalsize $^{2}${The Hong Kong University of Science
% and Technology (Guangzhou), Guangzhou, China}\\
% \normalsize $^{1}${\{guoyx, duzhuoran, kezhengtang, gcc, gaoyj\}@zju.edu.cn}~~~~~$^{2}${nantang@hkust-gz.edu.cn}\\
% }
% \renewcommand{\authors}{Yuxiang Guo, Zhuoran Du, Nan Tang, Kezheng Tang, Congcong Ge, and Yunjun Gao}
% \newcommand{\blue}[1]{\textcolor{blue}{#1}}

\author{Yuxiang Guo}
\affiliation{%
  \institution{Zhejiang University}
  \city{Hangzhou}
  \country{China}
  }
\email{guoyx@zju.edu.cn}
 
\author{Zhuoran Du}
\affiliation{%
  \institution{Zhejiang University}
  \city{Hangzhou}
  \country{China}
  }
\email{duzhuoran@zju.edu.cn}

\author{Nan Tang}
\affiliation{%
  \institution{The Hong Kong University of Science and Technology (Guangzhou)}
  \city{Guangzhou}
  \country{China}
  }
\email{nantang@hkust-gz.edu.cn}

\author{Kezheng Tang}
\affiliation{%
  \institution{Zhejiang University}
  \city{Hangzhou}
  \country{China}
  }
\email{kezhengtang@zju.edu.cn}

\author{Congcong Ge}
\affiliation{%
  \institution{Zhejiang University}
  \city{Hangzhou}
  \country{China}
  }
\email{gcc@zju.edu.cn}

\author{Yunjun Gao}
\affiliation{%
  \institution{Zhejiang University}
  \city{Hangzhou}
  \country{China}
  }
\email{gaoyj@zju.edu.cn}
\renewcommand{\shortauthors}{Yuxiang Guo et al.}

%% By default, the full list of authors will be used in the page
%% headers. Often, this list is too long, and will overlap
%% other information printed in the page headers. This command allows
%% the author to define a more concise list
%% of authors' names for this purpose.

%%
%% The abstract is a short summary of the work to be presented in the
%% article.
\begin{abstract}
Document-to-table (Doc2Table) extraction derives structured tables from unstructured documents under a target schema, enabling reliable and verifiable SQL-based data analytics. Although large language models (LLMs) have shown promise in flexible information extraction, their ability to produce precisely structured tables remains insufficiently understood, particularly for indirect extraction that requires complex capabilities such as reasoning and conflict resolution.  
Existing benchmarks neither explicitly distinguish nor comprehensively cover the diverse capabilities required in Doc2Table extraction. We argue that a capability-aware benchmark is essential for systematic evaluation. However, constructing such benchmarks using human-annotated document-table pairs is costly, difficult to scale, and limited in capability coverage. To address this, we adopt a reverse Table2Doc paradigm and design a multi-agent synthesis workflow to generate documents from ground-truth tables. Based on this approach, we present \textsf{DTBench}, a synthetic benchmark that adopts a proposed two-level taxonomy of Doc2Table capabilities, covering 5 major categories and 13 subcategories. We evaluate several mainstream LLMs on \textsf{DTBench}, and demonstrate substantial performance gaps across models, as well as persistent challenges in reasoning, faithfulness, and conflict resolution. \textsf{DTBench} provides a comprehensive testbed for data generation and evaluation, facilitating future research on Doc2Table extraction. The benchmark is publicly available at https://github.com/ZJU-DAILY/DTBench.

\end{abstract}

%% The code below is generated by the tool at http://dl.acm.org/ccs.cfm.
%% Please copy and paste the code instead of the example below.
%%



%%
%% Keywords. The author(s) should pick words that accurately describe
%% the work being presented. Separate the keywords with commas.

%% A "teaser" image appears between the author and affiliation
%% information and the body of the document, and typically spans the
%% page.

% \received{20 February 2007}
% \received[revised]{12 March 2009}
% \received[accepted]{5 June 2009}

%%
%% This command processes the author and affiliation and title
%% information and builds the first part of the formatted document.
\maketitle
\input{tex/1.Introduction}
\input{tex/2.Doc2Table}
\input{tex/3.Table2Doc}

\input{tex/4.Experiment}
\input{tex/5.RelatedWork}
\input{tex/6.Conclusion}
\input{tex/7.Acknowledgments}
\bibliographystyle{ACM-Reference-Format}
\bibliography{refer}
\input{tex/Appendix}
\end{document}

%% file: tex/1.Introduction.tex
\section{Introduction}
\label{sec:intro}

Modern organizations store vast amounts of unstructured data, such as clinical notes, legal contracts, and financial filings, accounting for more than 80\% of enterprise data~\cite{IDC}. Although Large Language Models (LLMs)~\cite{gpt42024,google2023gemini} have significantly advanced how organizations read and summarize documents, quantitative 
data analysis still requires structured data.
% LLMs cannot directly facilitate the SQL-like operations, such as aggregation, filter, and join, that are essential for quantitative decision-making.

% Unstructured data has become ubiquitous in the digital era, such as clinical notes, legal contracts, financial reports, and scientific literature, accounting for more than 80\% of enterprise data worldwide~\cite{IDC}. These unstructured data contain substantial yet underutilized business value. % Deriving valuable insights from unstructured data has emerged as one of the most critical challenges and opportunities in modern data analytics.

\down \noindent \textbf{Structured Tables Are Needed.} 
Consider a healthcare scenario in which medical institutions maintain large collections of unstructured documents, such as electronic health reports, clinical notes, and administrative records. A data analyst may pose a complex analytical query, e.g., ``\textit{What percentage of diabetic patients over 60 developed cardiovascular complications within six months of initiating Insulin Glargine?}''
Addressing such a query using LLMs or retrieval-augmented generation (RAG)~\cite{rag} is  unreliable. While RAG can retrieve isolated passages mentioning the drug or the complication, it cannot reliably aggregate evidence or perform deterministic computation~\cite{Hallucination}. Moreover, the resulting answers are opaque, preventing auditors from tracing which evidence contributes to the computation. In contrast, structured tables provide a more reliable foundation for accurate and verifiable analysis~\cite{Birdie, ARM-Net}. Consequently, a promising direction is to transform unstructured documents into structured tables, allowing queries to be answered through deterministic SQL execution~\cite{DataMosaic}.

\begin{figure*}
    % \vspace{-2mm}
	\centering
    \includegraphics[width=1\linewidth]{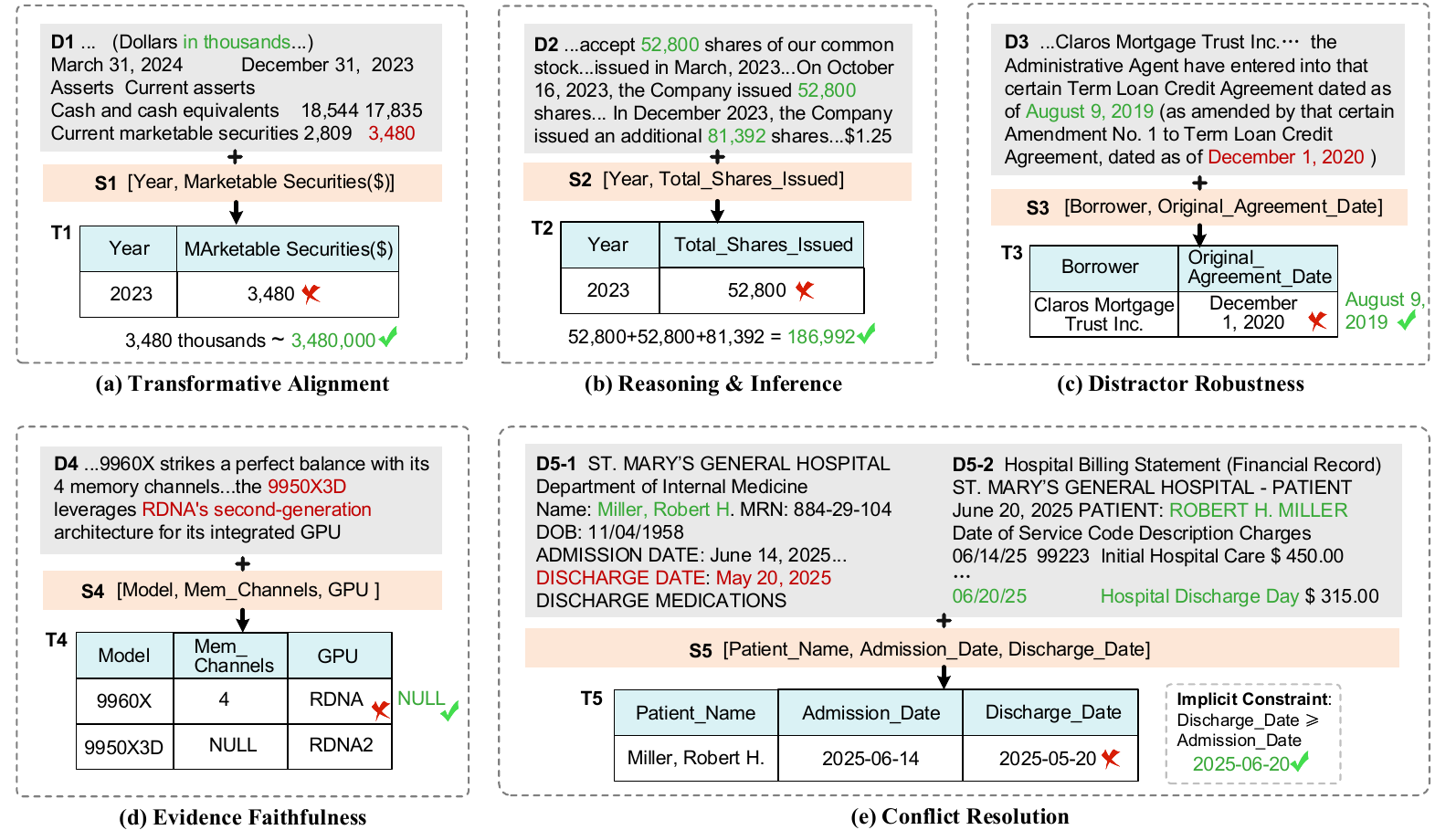}
    \vspace{-4mm}
    \caption{Challenging examples of Doc2Table extraction that require different capabilities.}
    \label{fig:example}
    \vspace{-2mm}
\end{figure*}

\down\noindent \textbf{Doc2Table Extraction.} 
Given the need for structured tables, recent studies~\cite{Doctopus,EVAPORATE,DocETL,Text2DB} employ LLMs to populate the target schema from source documents to produce a structured table. 
However, these systems largely abstract this document-to-table (Doc2Table) extraction as a generic LLM-based operation, without systematically examining its accuracy or reliability. 
In practice, Doc2Table is inherently non-trivial: many table values cannot be directly extracted from documents and instead require diverse capabilities beyond surface text matching. Figure~\ref{fig:example} presents representative error cases observed in real-world documents, 
illustrating different capabilities required to produce accurate and reliable tables.

\emp{(1) Transformative Alignment (TA).} 
When the target schema specifies a target value format or unit that differs from how the value is expressed in the source document, a transformation operation is required. As illustrated in Figure~\ref{fig:example}(a), the $(\$)$ notation in schema $S_1$ indicates unscaled U.S. dollar amounts, while the number 3,480 in the document is expressed in thousands. Therefore, the value should be transformed to obtain the correct amount 3,480,000.

\emp{(2) Reasoning and Inference (RI).}
Reasoning becomes necessary when the document does not explicitly state the target value, requiring logical inference or numerical computation to produce a schema-compliant result. As shown in Figure~\ref{fig:example}(b), the document reports three separate share issuances in 2023, while the schema $S_2$ requests the total value. Thus, it is necessary to identify and sum all three issuances to obtain the correct value of 186,992.

\emp{(3) Distractor Robustness (DR).}
This capability is required when the document contains semantically similar yet irrelevant information. In Figure~\ref{fig:example}(c), schema $S_3$ requires the \texttt{Original\_Agreement\_Date}. The document mentions both the original agreement date (August 9, 2019) and a subsequent amendment date (December 1, 2020). The extractor must successfully filter out the distracting amendment date to identify the correct original date.

\emp{(4) Evidence Faithfulness (EF).}
This capability ensures that the extracted table is strictly grounded in the source document. As shown in Figure~\ref{fig:example}(d), the text describes the GPU architecture for the ``9950X3D'' model but provides no GPU details for the ``9960X''. The extraction result should be ``NULL'', rather than incorrectly attributing the GPU features of the 9950X3D to the 9960X.

\emp{(5) Conflict Resolution (CR).}
When multiple sources or document segments provide conflicting information for the same attribute, conflict resolution is required to determine the correct value. As illustrated in Figure~\ref{fig:example}(e), document D5-1 lists a discharge date of ``May 20, 2025'', while document D5-2 lists ``June 20, 2025''. Given the admission date of ``June 14, 2025'', the implicit constraint that discharge cannot precede admission must be applied to resolve the conflict and select the correct value, 2025-06-20.

% \begin{table*}[t]
% \centering
% \small
% \vspace{-2mm}
% \caption{Comparison of existing benchmarks and our \textsf{DTBench} on five capability dimensions.}
% \vspace{-3mm}
% \label{tab:bench_comparison}
% \setlength{\tabcolsep}{6pt}
% \renewcommand{\arraystretch}{1}
% \begin{tabular}{l|cccccc}
% \toprule
% \textbf{Benchmark} &
% \makecell{\textbf{Transformative} \\ \textbf{Alignment (TA)}} &
% \makecell{\textbf{Reasoning \&} \\ \textbf{Inference (RI)}} & 
% \makecell{\textbf{Distractor} \\ \textbf{Robustness (DR)}} &
% \makecell{\textbf{Evidence} \\ \textbf{Faithfulness (EF)}} &
% \makecell{\textbf{Conflict} \\ \textbf{Resolution (CR)}} \\
% \midrule
% Rotowire~\cite{Rotowire}     & \cmark & \xmark & \xmark & \xmark & \xmark   \\
% E2E~\cite{E2E}  & \cmark & \xmark & \xmark & \xmark & \xmark  \\
% LiveSum~\cite{LiveSum}     & \xmark & \cmark & \cmark & \xmark & \xmark  \\
% InstructIE~\cite{InstructIE} & \xmark & \cmark & \cmark & \cmark & \xmark  \\
% StructText~\cite{StructText} & \cmark & \xmark & \cmark & \xmark & \xmark  \\
% \hline
% \textbf{\textsf{DTBench} (ours)}  & \cmark & \cmark & \cmark & \cmark & \cmark  \\
% \bottomrule
% \end{tabular}
% \vspace{-2mm}
% \end{table*}

\begin{table*}[t]
\centering
\small
\vspace{-2mm}
\caption{Comparison of existing benchmarks and our \textsf{DTBench} on five capability dimensions.}
\vspace{-3mm}
\label{tab:bench_comparison}
\setlength{\tabcolsep}{6pt}
\renewcommand{\arraystretch}{1} % 稍微增加行高，带颜色的表格看起来更美观
\begin{tabular}{l|ccccc} % 修正了列数，原代码里写了 6 个 c，实际只有 5 个能力维度
\toprule
\textbf{Benchmark} &
\makecell{\textbf{Transformative} \\ \textbf{Alignment (TA)}} &
\makecell{\textbf{Reasoning \&} \\ \textbf{Inference (RI)}} & 
\makecell{\textbf{Distractor} \\ \textbf{Robustness (DR)}} &
\makecell{\textbf{Evidence} \\ \textbf{Faithfulness (EF)}} &
\makecell{\textbf{Conflict} \\ \textbf{Resolution (CR)}} \\
\midrule
Rotowire~\cite{Rotowire}     & \cmark & \xmark & \xmark & \xmark & \xmark   \\
E2E~\cite{E2E}  & \cmark & \xmark & \xmark & \xmark & \xmark  \\
LiveSum~\cite{LiveSum}     & \xmark & \cmark & \cmark & \xmark & \xmark  \\
InstructIE~\cite{InstructIE} & \xmark & \cmark & \cmark & \cmark & \xmark  \\
StructText~\cite{StructText} & \cmark & \xmark & \cmark & \xmark & \xmark  \\
\hline
\rowcolor{gray!20} % 这里设置浅灰色背景 (20% 的灰色)
\textbf{\textsf{DTBench} (ours)}  & \cmark & \cmark & \cmark & \cmark & \cmark  \\
\bottomrule
\end{tabular}
% \vspace{-2mm}
\end{table*}

\noindent \textbf{The Need of a Benchmark Dataset.} It is reasonable to hypothesize that the capabilities discussed above vary substantially across LLMs. However, it remains unclear how different LLMs compare with respect to these capabilities. A comprehensive evaluation of Doc2Table is crucial, as extraction accuracy directly underpins downstream data analysis over structured tables. Unfortunately, existing benchmarks, e.g., Rotowrite~\cite{Rotowire}, E2E~\cite{E2E},  LiveSum~\cite{LiveSum}, InstructIE~\cite{InstructIE}, and StructText~\cite{StructText}, fail to cover the capabilities required for such an evaluation, as summarized in Table~\ref{tab:bench_comparison}. 
This motivates a comprehensive benchmark dataset.

\noindent \textbf{Challenges.} Developing a benchmark dataset that covers all five capability dimensions is non-trivial. 
Traditional methods require humans to collect documents, define schemata, read the documents, and manually extract structured tables as ground truth. This process is labor-intensive and inherently unscalable. More importantly, it offers limited control of task difficulty: deliberately constructing instances that target different capabilities typically requires extensive expert intervention and repeated dataset refinement, resulting in substantial additional cost.

\noindent \textbf{Our Proposal: Table2Doc Synthesis.} We propose a reverse approach: instead of manually annotating ground-truth tables from documents, we start with curated tables and synthesize documents that fully encode the table's information. This design ensures that ground-truth tables are inherently available, thus eliminating the prohibitive cost of manual annotation. Also, by carefully controlling the inverse generation, we can inject different difficulties into our generated dataset to evaluate different Doc2Table capabilities.

\down\noindent\textbf{Contributions.} Our main contributions are summarized as follows:
 
\noindent(1) \textbf{Doc2Table Capabilities Taxonomy}: We introduce a two-level taxonomy to classify the main capabilities required for Doc2Table extraction, to guide our benchmark design (Section~\ref{subsec:taxonomy}).

\noindent(2) \textbf{Table2Doc Synthesis Workflow}: We propose a 
document synthesis workflow that models capability as a latent variable to ensure coverage of the predefined capability taxonomy, and incorporates multi-stage verification to guarantee the quality of synthesized documents (Section~\ref{subsec:synthesis}).

\noindent(3) \textbf{New Benchmark}: We release \textsf{DTBench}, the first capability-aware benchmark for evaluating Doc2Table extraction. 
The dataset consists of 120 cases with 8,811 cell-level evaluation instances, covering all the capabilities defined in our taxonomy (Section~\ref{subsec:bench}).

\noindent(4) \textbf{Extensive Experiments}: We evaluate various mainstream LLMs on \textsf{DTBench} and conduct in-depth analyses to systematically assess their strengths and weaknesses, providing open research opportunities for Doc2Table extraction  (Section~\ref{sec:exp}).

% \down\noindent\textbf{Outline.}

%% file: tex/2.Doc2Table.tex
\section{Task Definition: Doc2Table Extraction}
In this section, we first provide the definition of document-to-table extraction, the task we aim to evaluate in this paper. Then, we propose a two-level taxonomy to categorize the crucial capabilities required for Doc2Table.

\subsection{Problem Statement}

In practice, enterprise documents are unstructured in form but structured in intent, as organizations operate under predefined schemata. This makes schema-guided document-to-table extraction both natural and necessary in real-world settings.
We first introduce the necessary notations and then formally define the task of Document-to-Table (Doc2Table) Extraction .

\vspace{0.5mm}\parhead{Target Schema ($S$)}: A normalized schema $S$ defines the structure of a table $T$ by specifying an entity type and the attribute set $A = \{a_1, a_2, \dots, a_m\}$. 
For each attribute, the schema may further specify metadata such as semantic descriptions, data types, value ranges, constraints, and representative examples.

\vspace{0.5mm}\parhead{Table ($T$)}: A structured table $T$ with schema $S$ is a collection of tuples $\{t_1, t_2, \dots, t_n\}$. Each tuple $t_i \in T$ refers to an entity, composed of a set of cells $\{t_i[a_1], t_i[a_2], \dots, t_i[a_m]\}$. Let $v_{ij} = t_i[a_j]$ be the value of $t_i$ on attribute $a_j$, and $\mathcal{V}= \{v_{ij}\}$ be the value domain of $T$.

\vspace{0.5mm}\parhead{Document ($D$)}: Let $D$ denote a document (e.g., a contract, exhibit, etc.) related to a single use case or business engagement. We assume that low-level parsing (e.g., layout detection and OCR~\cite{OCR}) has already been
performed; thus, this paper focuses on textual content. If multiple documents are provided, they can be concatenated into a single document.
\begin{myDef}
\textnormal{\textbf{(Doc2Table Extraction)}}.
Given a document $D$ and a target schema $S$, Doc2Table  Extraction  aims to seek a mapping 
$F: (D, S) \rightarrow T$ that produces a table $T$ consisting of $n$ tuples and $m$ attributes. Each tuple $t_i \in T$ contains $m$ cell values $\{v_{ij}\}_{j=1}^m$ that can be extracted directly or indirectly from $D$.
\end{myDef}

\subsection{Doc2Table Capabilities Taxonomy}
\label{subsec:taxonomy}
Note that the difficulty of extracting different cells $v_{ij}$ can vary substantially. For example, in Figure~\ref{fig:example}(a), ``2023'' is directly extractable, while ``17,835,000'' requires unit transformation. Motivated by such real-world cases and informed by extensive analyses of challenges in data extraction, cleaning, transformation, and fusion~\cite{DataMosaic,Camper,SourceDis}, we propose a comprehensive two-level taxonomy to categorize the capabilities required for Doc2Table extraction.

\parhead{Evidence ($E$)}: Given a document $D$ and a target schema $S$, let $T^*$ denote the ground-truth of Doc2Table. For each cell $v^*_{ij} \in T^*$, we define its evidence set $E_{ij}$ as a set of  text spans in $D$ that are relevant to the value $v^*_{ij}$. Let $\mathcal{E}$ denote the set of all evidence sets associated with the cell values in the value domain $\mathcal{V^*}$ of $T^*$. 
% , and $\mathcal{V}$ denote the domain of table values.

\begin{myDef}
\textnormal{\textbf{(Doc2Table Capability)}}. 
A Doc2Table capability $c$ specifies the functional logic required to correctly determine a table cell value from its evidence set.
Formally, each capability $c$ induces a mapping $\phi_c: \mathcal{E} \to \mathcal{V}^*$, such that $\phi_c(E_{ij}) = v^*_{ij}$, where $E_{ij} \in \mathcal{E}$ is the evidence set associated with ground-truth cell $v^*_{ij}$.
\end{myDef}

\begin{figure*}
	\centering
    % \vspace{-5mm}
    \includegraphics[width=1\linewidth]{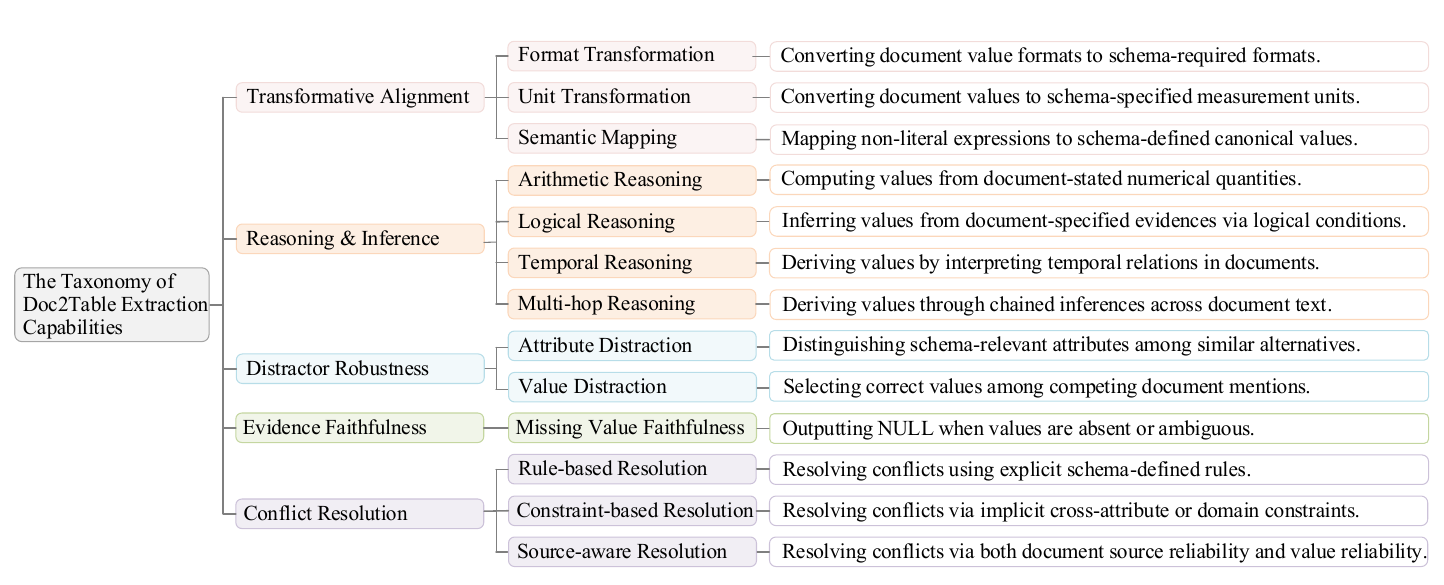}
    \vspace{-8mm}
    \caption{A Two-level Taxonomy of Doc2Table Extraction Capabilities.}
    \label{fig:taxonomy}
    \vspace{-2mm}
\end{figure*}

The required Doc2Table capabilities naturally exhibit a hierarchical structure, comprising coarse-grained functional categories (e.g., Transformative Alignment) and finer-grained sub-capabilities (e.g., Format Transformation, Unit Transformation, etc.). Accordingly, we organize them into five top-level categories, most of which are further divided into multiple sub-capabilities, as shown in Figure~\ref{fig:taxonomy}. We denote the resulting two-level capability taxonomy as $\mathcal{H}$, where each top-level capability $c_k \in \mathcal{H}$ consists of a set of second-level sub-capabilities ${c_{k,l}}$.

\subsubsection{Transformative Alignment (TA)}
This capability is required when a value appears in the source document but does not conform to the constraints of schema $S$. In such cases, the value must be transformed into a semantically equivalent form that satisfies the schema without changing its underlying meaning. This capability is further divided into three sub-capabilities.
 
\parhead{Format Transformation.}
The schema $S$ enforces a specific attribute format (e.g.,  ``yyyy-mm-dd''), while the document uses an alternative representation, necessitating format transformation.

\parhead{Unit Transformation.}
The schema $S$ requires a specific unit (e.g., kilometers) for a numerical attribute, while the document uses a different unit (e.g., meters), necessitating conversion.

\parhead{Semantic Mapping.}
The schema $S$ specifies an abstract or canonical representation for an attribute (e.g., a numeric rank), while the corresponding value in the document is expressed using a semantically related but non-equivalent concept (e.g., ``silver medal''), requiring semantic mapping to convert the  expression into the schema-defined representation.

% \parhead{\blue{Abbreviation Normalization.}}
% The schema $S$ specifies a canonical abbreviated form for a categorical attribute value (e.g., ``NY''), while the corresponding value in the source document may appear in its full or synonymous form (e.g., ``New York''), requiring normalization to the schema-defined representation.

\subsubsection{Reasoning \& Inference (RI)}
This capability is required when the target table value is not explicitly stated in the document, but must be derived from indirect, partial, or scattered textual evidences to satisfy the schema requirements. We categorize the reasoning into four subcategories.

\parhead{Arithmetic Reasoning.}
The document provides relevant numerical quantities, but the specific attribute requires a derived value that must be computed from them, rather than directly extracted.

\parhead{Logical Reasoning.}
The target value depends on logical conditions or constraints described in the document, requiring inference beyond surface-level value matching.

\parhead{Temporal Reasoning.}
The document describes events or values over time, requiring inference from temporal relations instead of direct extraction of a single time expression.

\parhead{Multi-hop Reasoning.}
The evidences supporting a target attribute 
is scattered across multiple parts of the document, requiring multi-step reasoning and computation.

\subsubsection{Distractor Robustness (DR)}
This capability is required when misleading, irrelevant, or confusing information in the source document challenges the correct extraction of a target value. Such distractions arise from two main types:

\parhead{Attribute Distraction.}
The document contains multiple attributes with similar semantics or forms, increasing the risk of selecting a value associated with an incorrect attribute.

\parhead{Value Distraction.}
Multiple candidate values are mentioned for the same attribute, including outdated, erroneous, or superseded information, requiring careful identification of the correct value.

\subsubsection{Evidence Faithfulness (EF)}
\label{subsubsec:EF}
Evidence Faithfulness requires that any extracted or inferred value be strictly grounded in the source document. A value is admissible only if it can be uniquely derived from explicit textual evidence using deterministic rules or widely accepted conventions. Inferences based solely on a model's internal knowledge without textual support are considered hallucinations and are disallowed.

\parhead{Missing Value Faithfulness.}
``\textsf{NULL}'' should be populated when a target value is missing or not uniquely inferable from the document; generating a plausible value in such cases is considered unfaithful.

\subsubsection{Conflict Resolution (CR)}
\label{subsubsec:CR}
This capability is required when multiple conflicting candidate values for the same attribute and entity appear in the source document(s). In such cases, the correct value must be selected according to principled resolution criteria. This capability comprises three forms.

\parhead{Rule-based Resolution.}
Conflicts can be resolved by applying explicit resolution rules specified by the schema $S$, such as prioritizing more recent timestamps or higher-precision measurements.

\parhead{Constraint-based Resolution.}
Conflicts are resolved by leveraging implicit cross-attribute constraints or domain consistency requirements, enabling the correct value to be inferred even in the absence of explicit resolution rules (e.g., Figure~\ref{fig:example}(e)).

\parhead{Source-aware Resolution.}
When neither explicit resolution rules nor implicit constraints are available, conflicts must be resolved through joint reasoning over document source reliability and candidate value reliability to infer the most plausible value, a process commonly referred to as data fusion~\cite{SourceDis}. For example, in stock reporting, the same attribute may be reported by multiple sources with inconsistent or conflicting values.

%% file: tex/3.Table2Doc.tex
\section{Dataset Construction: Table2Doc Synthesis}

\begin{figure*}
	% \centering
 %    \vspace{-2mm}
    \includegraphics[width=1\linewidth]{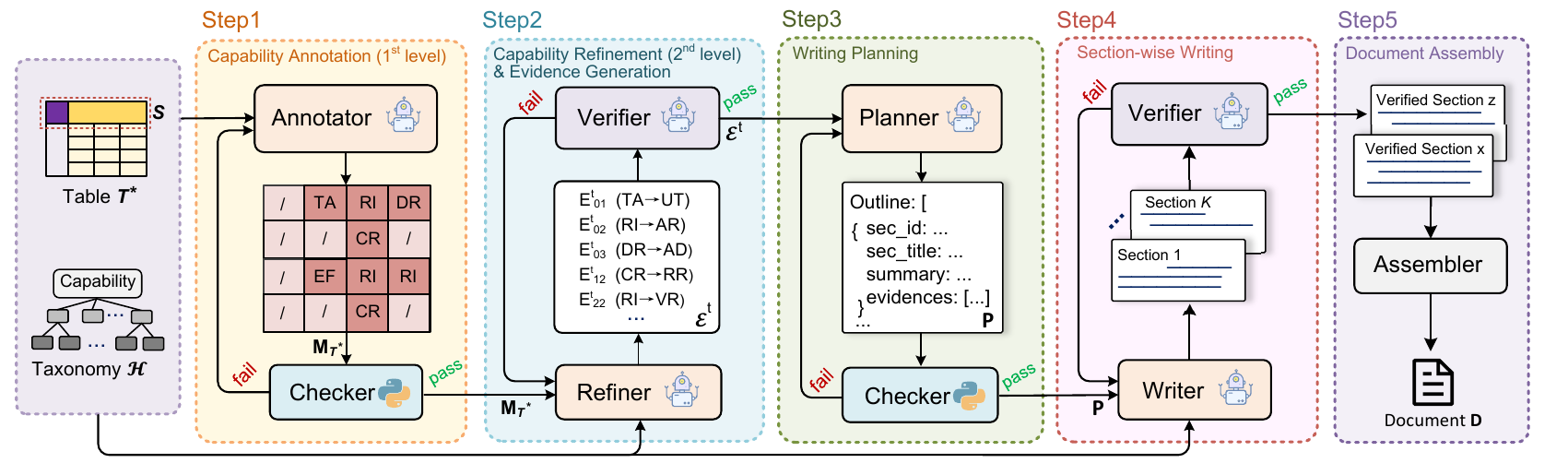}
    \vspace{-5mm}
    \caption{Overview of the proposed multi-agent workflow for Table2Doc synthesis.}
    \label{fig:overview}
    % \vspace{-2mm}
\end{figure*}

In this section, we first formalize the Table2Doc synthesis problem for benchmark construction. We then analyze the key challenges of Table2Doc, followed by the introduction of a multi-agent workflow for Table2Doc synthesis. Finally, we summarize the characteristics of the resulting benchmark dataset, \textsf{DTBench}.

\subsection{Definition}
To construct a benchmark for evaluating Doc2Table extraction, we adopt a \textbf{reverse perspective}: starting from a ground-truth table $T^*$ with the schema $S$, we aim to synthesize a document $D$ that can be extracted into the table $T^*$ given the target schema $S$. We formally define the Table2Doc Synthesis as follows.

\begin{myDef}
\label{def:table2doc}
\textnormal{\textbf{(Table2Doc Synthesis)}.}
Given a table $T^*$, a schema $S$, and a two-level capability taxonomy $\mathcal{H}$,
Table2Doc aims to synthesize a text document $D$ such that:
(i)  Capability Awareness: extracting $T^*$ from $D$ necessitates the application of specific capabilities defined in $\mathcal{H}$;
(ii) Completeness: each cell value $v^*_{ij} \in T^*$ can be extracted from the document $D$; and
(iii) Exclusiveness: no table $T \supset T^*$ can be extracted from $D$, i.e., $D$ does not contain any additional tuples extractable under the given schema $S$ beyond those in $T^*$.

\end{myDef}

% After synthesis, each test case consists of a triple $(D, S, T^*)$, which serves as an evaluation instance for the Doc2Table extraction. 
Properties (ii) and (iii) guarantee that $T^*$ is the unique ground-truth table recoverable from the document $D$ under the schema $S$, while property (i) enables fine-grained, cell-level evaluation of distinct Doc2Table extraction capabilities.
We do not require each case to cover all capability categories in the taxonomy $\mathcal{H}$, as enforcing full coverage would be unnatural in practice. For example, when a table $T^*$ contains no numerical attributes, requiring numerical computation during synthesis is inappropriate.
% Instead, each synthesized document $D$ is designed to remain natural and logically coherent, closely resembling realistic real-world documents rather than artificial constructions.

\subsection{Challenges \& Overview}
% Synthesizing a document $D$ while satisfying the properties defined above is a non-trivial task. 
There are two main challenges for Table2Doc synthesis:

\noindent \textbf{(C1) Inverse Generation of Capability-Aware Evidence.}
In standard Doc2Table extraction, capabilities are identified by jointly observing the document evidence and the resulting table values. 
In contrast, Table2Doc synthesis requires generating evidences reversely based on a given table value, i.e., modeling $\mathbb{P}(E_{ij} \mid v_{ij})$.
This distribution is ill-posed, since the same value can be supported by diverse evidences corresponding to different capabilities. The capabilities, however, are not explicitly specified during reverse synthesis and may vary across multiple valid evidence realizations, leading to a circular dependency between capability selection and evidence generation.

To break this circularity, we leverage the observation that many capabilities are partially inferable from the cell value and schema context alone, and treat capability as a \textit{latent variable} during synthesis. Specifically, we first predict a coarse capability $c_k \sim \mathbb{P}(c_k \mid v_{ij})$, and then refine it to a sub-capability $c_{k,l} \sim \mathbb{P}(c_{k,l}  \mid v_{ij}, c_k )$, enabling evidence generation conditioned on both the cell value and the concrete sub-capability, i.e., $\mathbb{P}(E_{ij} \mid v_{ij}, c_{k,l} )$.
% To address this challenge, we factorize the ill-posed distribution $\mathbb{P}(E_{ij} \mid v_{ij})$ by \textit{introducing capability as a latent variable}.
% An LLM first predicts a coarse-grained first-layer capability $c_k \sim \mathbb{P}(c_k \mid v_{ij})$ from the cell value and schema context, constraining the evidence space. A refinement agent then selects a second-layer sub-capability $c_{k,l}  \sim \mathbb{P}(c_{k,l} \mid v_{ij}, c_k)$ and generates evidence conditioned on both the value and sub-capability, i.e., $\mathbb{P}(E_{ij} \mid v_{ij}, c_{k,l})$. This hierarchical factorization breaks the circular dependency and enables  

% decouple capability assignment from evidence generation.
% First, we leverage schema and cell-level context to predict a coarse-grained capability from the first layer of the taxonomy, breaking the circular dependency between observing evidence and selecting capabilities.
% Second, conditioned on the assigned capability, a refiner agent determines a concrete sub-capability and performs reverse evidence generation, enabling controlled and well-specified synthesis.

\noindent\textbf{(C2) Ensuring Completeness and Exclusiveness.}
Another key challenge in Table2Doc synthesis is to guarantee both \emph{completeness} and \emph{exclusiveness} of the synthesized document with respect to the ground-truth table $T^*$.  Violating either property compromises the uniqueness of the ground-truth table.
This requirement fundamentally differs from conventional LLM-based text generation~\cite{Recurrentgpt,Ex3,StructText}, which prioritizes fluency and plausibility. In contrast, Table2Doc synthesis requires fine-grained control and systematic verification to enforce both properties. However, the unstructured and lengthy nature of free-form text makes controllable generation and post-hoc verification particularly challenging. 

To address this, we adopt two complementary strategies. 
First, we introduce structure-guided generation to improve controllability,   using explicit schema constraints and evidence-based writing plans to guide the synthesis process.
Second, we prioritize deterministic verification whenever possible and employ checklist-based validation for cases requiring LLM-based judgment, thereby improving verification reliability.

\down\noindent\textbf{Overview.}
We propose a multi-agent workflow for Table2Doc synthesis, consisting of five sequential stages, as illustrated in Figure~\ref{fig:overview}.
% The key insight is that, by explicitly orchestrating the document generation process, we can systematically inject targeted Doc2Table extraction capabilities into synthetic documents while preserving semantic consistency with the ground-truth table.
Given a table $T^*$ and a predefined two-level capability taxonomy $\mathcal{H}$, Step~1 assigns first-layer capability labels to each cell in $T^*$, resulting in a capability matrix $\mathbf{M}_{T^*}$. Step 2 refines $\mathbf{M}_{T^*}$ by selecting second-layer sub-capabilities and generates structured evidences $\mathcal{E}^t$ conditioned on the cell values and assigned sub-capabilities. These evidence items are iteratively validated using a checklist-based verifier until all criteria are satisfied.
Step 3 constructs a document-level writing plan $\mathbf{P}$ from the verified evidences, which is then validated using deterministic code. 
Step 4 performs section-wise document generation guided by $\mathbf{P}$, with each section verified by a checklist-based verifier and revised if necessary.
Finally, Step 5 assembles all verified sections into a resulting document $D$.

\subsection{Multi-Agent Workflow}
\label{subsec:synthesis}

\subsubsection{Step 1: Capability Annotation}
Given a curated table  $T^*$ with schema $S$ and the capability taxonomy $\mathcal{H}$, we prompt an LLM-based \textsf{Annotator} to understand the input table, and assign each cell either a proper capability label $c_k \in \mathcal{H}$ or ``empty'' label based on the cell value and its schema context, yielding a capability matrix $\mathbf{M}_{T^{*}}$. We allow a cell to be assigned an ``empty'' label, indicating that its value is explicitly stated in the source document and can be directly extracted. Although this annotation process could be performed manually (e.g., by assigning all numerical values to a fixed reasoning capability), we adopt an LLM to approximate the latent distribution $\mathbb{P}(c_k \mid v_{ij})$. The stochasticity of LLM-driven annotation yields more diverse and heterogeneous capability assignments, better capturing the variability of real-world documents.

A deterministic, code-based \textsf{Checker} is then employed to verify the completeness of $\mathbf{M}_{T^*}$. Any unannotated cells are iteratively returned to the annotator for annotation. After three rounds, any remaining unannotated cells are assigned the ``empty'' label.

\subsubsection{Step 2: Refinement \& Evidence Generation}
Given a capability matrix $\mathbf{M}_{T^*}$, the LLM-based \textsf{Refiner} first assigns a concrete sub-capability $c_{k,l}$ to each cell whose label $c_k$ is not empty.
Then, the \textsf{Refiner} performs inverse evidence synthesis. 
For each cell value $v_{ij}$, we first construct a canonical form of its original evidence. Specifically, we represent $v_{ij}$ as a structured triple and verbalize it into a textual evidence $e^o_{ij}$ in the form: ``the attribute $a_j$ of entity $v_{i1}$ is $v_{ij}$'', where $v_{i1}$ denotes the entity name of $i$-th tuple (see Appendix~\ref{app:data_collect}).
Given an assigned sub-capability, the original evidence $e^o_{ij}$ is then inversely transformed into a capability-specific evidence $e^t_{ij}$ according to the semantics of the sub-capability. This inverse transformation is strictly constrained by the target schema $S$: the generated evidence must not introduce any additional schema-conformable facts. This constraint prevents the generation of plausible but extraneous facts that could be erroneously mapped to other attributes and extracted as spurious tuples.
The transformed evidence $e^t_{ij}$ is further decomposed into multiple fragments when it contains multiple sub-evidences, forming a reverse evidence set $E^t_{ij}$. This decomposition facilitates flexible section organization in the subsequent stage. 

% \begin{example}
%     Consider a cell with a value of 10 million, which is assigned to the Arithmetic Reasoning capability in step~1, indicating original evidence $e^o$: ``Company A's total profit in the last year is 10 million''. In step~2, the refiner first assigns the Basic Arithmetic sub-capability, and then performs inverse evidence synthesis by rewriting the original evidence $e^o$ into $e^t$: ``Company A's profit in this year is 8 million, which is 2 million less than last year''. Since $e^t$ is compositional, it can be further decomposed into two independent pieces of evidence: $e^{t,1}$: ``Company A's profit in this year is 8 million''; and $e^{t,2}$: ``Company A's profit decreased by 2 million compared to last year'', forming an reverse evidence set $E^t =\{e^{t,1}, e^{t,2}\}$.  
% \end{example}

Since inverse evidence synthesis relies on LLM-based generation, it requires careful verification. We employ an LLM-based \textsf{Verifier} to evaluate each reverse evidence set $E^t_{ij}$ using a checklist with three dimensions:
(i) \textit{value correctness}: whether the inverse evidence set $E^t_{ij}$ faithfully support the recovery of the original cell value $v_{ij}$; (ii) \textit{label alignment}: whether the recovery process requires the specified sub-capability $c_{k,l}$; and
(iii) \textit{schema leakage}: whether the evidence set contains any extractable schema-compatible facts beyond $v_{ij}$.
If verification fails, the verifier returns structured feedback, including a failure rationale identifying the inconsistency and a revision suggestion. This feedback is appended to the context to guide the refiner in regenerating the evidence set. The process iterates until verification succeeds or a predefined retry limit is reached.

\subsubsection{Step 3: Section Planning} Given the set of reverse evidences $\mathcal{E}^t = \bigcup_{i=1}^{n} \bigcup_{j=1}^{m} { E^t_{ij} }$, we employ an LLM-based \textsf{Planner} to construct a writing plan.
The planner first determines an appropriate document type (e.g., a financial report, a clinical trial summary, etc.) based on the content of $\mathcal{E}^t$. It then generates a structured document blueprint $\mathbf{P}$ comprising $K$ sections. Each section $s_k$ is associated with a title, a brief summary, and a subset of evidences $E^t_k \subseteq \mathcal{E}^t$ to be covered. Evidences are flexibly assigned to sections according to semantic relevance and document coherence.

A deterministic, code-based \textsf{Checker} then verifies whether all evidences are covered by the plan $\mathbf{P}$. If any evidence is missing, the planner is triggered to revise $\mathbf{P}$ by inserting the omitted evidences into appropriate sections, ensuring complete coverage before document generation.

\subsubsection{Step 4: Section-wise Writing}
In this step,  we employ an LLM-based \textsf{Writer} to generate text based on the given schema $S$ and the writing plan $\mathbf{P}$. Since LLMs struggle to produce long text in a single pass~\cite{longwriter}, writing proceeds section by section. Similar to step~2, the schema $S$ is used to prevent the writer from introducing schema-conformable facts beyond those explicitly listed in the plan.
% To ensure traceability, the writer is required to wrap all table-related statements in special markers, enabling more reliable verification by either an LLM-based verifier or human reviewers.

We further employ an LLM-based \textsf{Verifier} to evaluate each generated section using a two-dimensional checklist: (i) \textit{faithful grounding}: whether all planned evidences are expressed in the section and are extractable from the text; and (ii) \textit{schema leakage}: whether any schema-conformable content not grounded in the ground-truth table is present, thereby preventing the inclusion of plausible but unsupported facts.
Sections that fail verification, together with the corresponding error reasons, are fed back to the writer for targeted revision and re-evaluated in an iterative loop. Sections that pass verification are forwarded to the subsequent assembly stage to construct the final document.

\subsubsection{Step 5: Document Assembly}
Finally, all sections validated in the previous stage are gathered by the \textsf{Assembler}. This module performs the final structural integration, synthesizing these discrete segments into a single document $D$ and completes the Table2Doc synthesis pipeline.

\begin{figure}[t]
    \centering

    \captionof{table}{Statistics of documents and tables in \textsf{DTBench}.}
    \label{tab:dtbench_stats}
    \vspace{-3mm}
    \small
    \setlength{\tabcolsep}{4pt} 
    \begin{tabular}{p{3.5cm}ccc} 
        \toprule
        \textbf{Attribute} & \textbf{Min} & \textbf{Max} & \textbf{Avg} \\
        \midrule
        % \multicolumn{4}{l}{\textbf{Document ($D$)}} \\ 
         \#Row of  $T^*$    & 3 & 37 & 12.2 \\
         \#Column of  $T^*$ & 2 & 17 & 7.2 \\
        
        \midrule 
        % \multicolumn{4}{l}{\textbf{Ground-Truth Table ($T^*$)}} \\
        Length (\#Tokens) of $D$ & 498 & 182,297 & 28,563 \\   
        \bottomrule
    \end{tabular}

    \vspace{1mm} % 

    % --- 第二部分：图片 ---
    \includegraphics[width=1\linewidth]{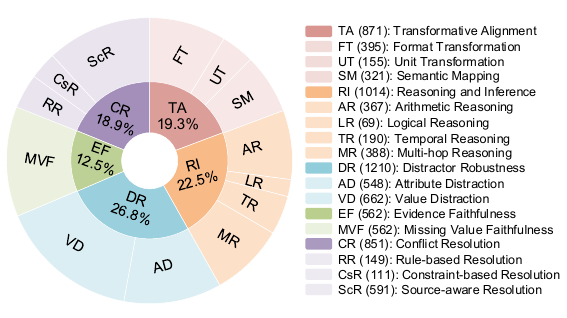}
    % \vspace{-3mm}
    \captionof{figure}{Proportion of cells annotated with different capabilities and sub-capabilities in \textsf{DTBench}.}
    \vspace{-6mm}
    \label{fig:dtbench_dist}
    
\end{figure}

% \down\noindent\textbf{Advantages.} Our multi-agent framework offers several key advantages for benchmark construction: (1) Controllability: By separating planning, refinement, and writing, we can systematically control both contextual complexity (through section organization) and operational complexity (through strategy assignment) at appropriate granularities. (2) Quality Assurance: The integrated verification mechanisms at both refinement and writing stages ensure semantic consistency and prevent common synthesis pitfalls like information omission or hallucination. (3) Scalability: The modular design with parallel processing and caching enables efficient synthesis of large-scale benchmarks. (4) Interpretability: Each synthesis stage produces interpretable intermediate outputs (plans, refined instructions), facilitating debugging and quality analysis.

\subsection{\textsf{DTBench}}
\label{subsec:bench}
We apply the proposed workflow to construct the benchmark dataset \textsf{DTBench}.
% We collect real-world tables as ground-truth $T^*$, and apply the proposed multi-agent workflow to construct a benchmark \textsf{DTBench}. 
Specifically, we curate 120 tables $T^*$ from Kaggle~\cite{kaggle}, Wikipedia~\cite{wikipedia}, and public data fusion datasets~\cite{fusiondataset}, spanning diverse domains, each containing an entity identifier column.
To enable controlled evaluation of faithfulness, we deliberately remove a subset of cell values and treat them as missing. 
We further manually augment the schema $S$ with attribute descriptions, data types, and constraints. 
Tables from data fusion datasets~\cite{fusiondataset} are included to inject source-aware resolution capability, as these datasets contain a gold table and multiple conflicting facts collected from different data sources. 
Further details of the table collection and processing are provided in Appendix~\ref{app:data_collect}. We use Grok-4-fast~\cite{grok4fast} as the backbone for all agents in our document synthesis workflow, based on preliminary experiments that consider both generation quality and cost. We also evaluate the quality of the generated documents along three dimensions, as detailed in Appendix~\ref{app:doc_eva}.

\textsf{DTBench} comprises 120 synthesized cases $(D, S, T^*)$ with 8,811 cell-level instances, of which 4,518 (over 50\%) are annotated with capability labels from our taxonomy. 
Table~\ref{tab:dtbench_stats} summarizes key statistics of the ground-truth tables $T^*$ and the synthesized documents $D$. 
Figure~\ref{fig:dtbench_dist} shows the distribution of cells across different capability categories. Cells without capability annotations correspond to directly extractable cells.
% , which require only replication from the source document.

% \vspace{1mm}
\noindent \textbf{Discussion.}
While model bias is an inherent challenge in LLM-synthesized datasets, we mitigate this through the following core design:
(i) Task Complexity: \textsf{DTBench} evaluates the Doc2Table extraction task, which requires complex, specific capabilities (e.g., multi-step reasoning and conflict resolution) rather than surface-level text matching. Consequently, evaluated LLMs cannot gain an unfair advantage simply by aligning with the synthesizer's writing style.
(ii) Constrained Synthesis: Rather than unconstrained free generation, our multi-agent data synthesis pipeline is strictly guided by structured evidence and writing plans, effectively mitigating the synthesizer's stylistic footprint.
Furthermore, to demonstrate that our benchmark evaluation and key findings are not affected by model bias, 
we reconstruct the complete dataset using GPT-4o-mini as an alternative synthesizer to Grok-4-fast. The detailed evaluation results are provided in Appendix~\ref{app:new_DTBench}.

%% file: tex/4.Experiment.tex
\section{Experiments}
\label{sec:exp}
In this section, we demonstrate the effectiveness of our proposed document synthesis workflow, and evaluate different LLMs for Doc2Table extraction on \textsf{DTBench}.
Specifically, we try to answer the following questions.

\noindent \textbf{Q1:} How effective are the key components of our multi-agent workflow for Table2Doc synthesis?

\noindent \textbf{Q2:} How effective are different LLMs at Doc2Table extraction? 

\noindent \textbf{Q3:} What are the strengths and weaknesses of these LLMs across different capabilities defined in our taxonomy? 
% \blue{We further provide a counterfactual validation experiment in 
% Appendix~\ref{app:counter_val} to assess the potential impact of parametric knowledge on extraction accuracy.}

\subsection{Setting}

\noindent\textbf{Models.} We consider LLMs of different families and scales, including Qwen3-4B~\cite{qwen3}, Qwen3-32B~\cite{qwen3}, Llama3.1-8B~\cite{llama3}, Llama3.1-70B~\cite{llama3},  Deepseek-V3.2~\cite{DeepSeekV3.2}, GPT-5-mini~\cite{gpt5mini}, GPT-5~\cite{gpt5}, and Gemini-3-flash~\cite{gemini3flash}. For all models, we set the temperature to 0 to ensure deterministic output.
Although some Table Foundation Models have been proposed recently~\cite{tablegpt, tablellm},
these models are typically optimized for tasks where the input is a structured table. In contrast, \textsf{DTBench} evaluates Doc2Table extraction, where the input is an unstructured document. Due to this task mismatch, Table foundation models are often ill-suited for document extraction. To verify this, we conduct an additional experiment using a recent TableLLM~\cite{tablellm} (see appendix~\ref{app:tablellm}).

% \noindent\textbf{Dataset.} Experiments are conducted on  \textsf{DTBench}.

\noindent\textbf{Cell Alignment for Evaluation.}
Given an input document $D$ and target schema $S$, we compare model predicted table $T$ and the ground-truth table $T^*$ at the cell level.  
Since the schema $S$ is predefined, column alignment between $T$ and $T^*$ is straightforward, requiring no additional processing.
However, the number and ordering of tuples in $T$ and $T^*$ may differ, rendering direct row-wise comparison infeasible.
To address this, we formulate tuple alignment as a weighted bipartite matching problem based on key attribute similarity, enforcing a one-to-one correspondence between matched rows (details in Appendix~\ref{app:alignment}).
This process yields a set of aligned cell pairs between $T$ and $T^*$.

\begin{table}[t]
\centering
\small
\caption{Ablation study of multi-agent Table2Doc synthesis.}
\label{tab:ablation}
\vspace{-3mm}
\setlength{\tabcolsep}{6.2pt}  
\label{tab:ablation_three_line}
\begin{tabular}{l | c c c c c c}
\toprule
\textbf{Variants} & \textbf{LR} & \textbf{LC} & \textbf{TC} & \textbf{CA(\%)} & \textbf{CP(\%)} & \textbf{EX(\%)} \\
\midrule
\textbf{Ours}     & \textbf{4.85} & \textbf{4.35} & \textbf{4.10} & \textbf{58.90} & \textbf{100} & \textbf{100} \\
\hline
w/o \textsf{Annotator}     & 4.40          & 4.15          & 3.95          & 1.53           & 98.17        & 95.16        \\
w/o \textsf{Refiner}       & 4.65          & 4.10          & 3.95          & 58.70          & 99.69        & 99.47        \\
w/o \textsf{Planner}       & 4.25          & 3.85          & 3.45          & 39.17          & 99.90        & 98.59        \\
w/o \textsf{Verifier}      & \textbf{4.85} & 4.20          & 4.00          & 28.79          & 72.43        & 88.73        \\
\bottomrule
\end{tabular}
\vspace{-4mm}
\end{table}
\subsection{Experiment for Q1}
We conduct an ablation study to demonstrate the necessity of each agent. Specifically, we systematically removed each of the agents \textsf{Annotator}, \textsf{Refiner}, \textsf{Planner}, and \textsf{Verifier} from our Table2Doc synthesis pipeline one by one, creating four variant document synthesis pipelines. The \textsf{Writer} is retained because it is the foundational engine; without it, no document can be synthesized for evaluation. Then, we randomly sample 50 tables from our collected table corpus and use the four variant pipelines to generate documents, and compare them with the documents in \textsf{DTBench}.

\vspace{1mm}
\noindent \textbf{Metrics.} We employ DeepSeek-V3.2 as a judge to evaluate the documents using both linguistic metrics (scored on a 1-5 scale) and task-specific metrics (measured as percentages, 0–100\%). The linguistic metrics encompass Lexical Richness (LR), Logical Consistency (LC), and Textual Coherence (TC), as defined in Appendix~\ref{app:doc_eva}. The task-specific metrics (see Definition~\ref{def:table2doc}) include: Capability-awareness (CA), defined as the proportion of cells requiring specific capabilities to be extracted; Completeness (CP), the proportion of cells successfully recovered from the generated document; and Exclusiveness (EX), which is 1 minus the proportion of hallucinated cells (i.e., extra cells extracted due to schema leakage that are absent from the ground-truth table).

Table~\ref{tab:ablation} presents the results. We have the following observations:
(i) The \textsf{Annotator} primarily dictates the difficulty of the dataset. Without it, most cells become directly extractable.
(ii) The \textsf{Refiner} and Planner collaboratively improve the linguistic quality (e.g., readability and coherence) of the generated documents.
(iii) The \textsf{Verifier}s are important to the correctness of the synthesized datasets, contributing to high Completeness (CP) and Exclusiveness (EX).

\subsection{Experiment for Q2}
\noindent\textbf{Metrics.}
We evaluate LLMs on all test cases using Precision ($P$), Recall ($R$), and F1. After aligning predicted and ground-truth cells, we apply standard normalization (e.g., whitespace and case normalization, punctuation removal) and count exact matches. Let $TP_i$ denote the set of matched cell pairs for test case $i$. The metrics are defined as: 
% $P =  \sum_{i=1}^N |TP_i|/\sum_{i=1}^N |C_{T_i}|$, $R =  \sum_{i=1}^N |TP_i| / \sum_{i=1}^N |C_{T_i^*}|$, $F1 =  2PR / (P+R)$,
\[
P = \frac{\sum_{i=1}^N |TP_i|}{\sum_{i=1}^N |C_{T_i}|}, \quad
R = \frac{\sum_{i=1}^N |TP_i|}{\sum_{i=1}^N |C_{T_i^*}|}, \quad
F1 = \frac{2PR}{P+R}
\]
where  $|C_{T_i}|$ and $|C_{T_i^*}|$ represent the total number of cells in the predicted and ground-truth tables, respectively.

To distinguish direct and indirect extraction, we partition the ground-truth cells $C_{T_i^*}$ into capability-demanding cells $C^{\text{ind}}_{T^*}$ and directly extractable cells $C^{\text{dir}}_{T^*}$, according to whether a capability label is assigned. We then report type-specific recall: 
% $R_{\text{ind}} =  \sum_{i=1}^N |TP_i \cap C^{\text{ind}}_{T^*}|/\sum_{i=1}^N |C^{\text{ind}}_{T^*}|$, $R_{\text{dir}} =  \sum_{i=1}^N |TP_i \cap C^{\text{dir}}_{T^*}|/\sum_{i=1}^N |C^{\text{dir}}_{T^*}|$.
\[
R_{\text{ind}} = \frac{\sum_{i=1}^N |TP_i \cap C^{\text{ind}}_{T^*}|}{\sum_{i=1}^N |C^{\text{ind}}_{T^*}|}, \quad
R_{\text{dir}} = \frac{\sum_{i=1}^N |TP_i \cap C^{\text{dir}}_{T^*}|}{\sum_{i=1}^N |C^{\text{dir}}_{T^*}|}.
\]
Note that we report a unified Precision, as non-aligned predicted cells cannot be reliably categorized as direct or indirect errors.

We also report the average Time (in seconds) per document and the total Cost (in USD) for different LLMs on \textsf{DTBench}.

\begin{table}[t]
\centering
\footnotesize
\caption{Overall performance of LLMs on \textsf{DTBench}.}
\label{tab:overall_performance}
\vspace{-3mm}
\renewcommand{\arraystretch}{0.9} 
\setlength{\tabcolsep}{3.1pt}  
\begin{tabular}{l | c c c c c  >{\columncolor{gray!20}}c c c}
\toprule
\textbf{Models} & $\bm{P}$ & $\bm{R}$ & $\bm{F1}$ & $\bm{R}_\text{dir}$ & $\bm{R}_\text{ind}$ & \multicolumn{1}{>{\columncolor{gray!20}}c}{$\bm{\Delta}_{\text{dir}}^{\text{ind}}$ } & Time & Cost \\  
\hline
\textbf{Qwen3-4B}       & 44.09 & 27.20 & 33.65  & 35.47 & 19.33 & $\downarrow$ 45.50\% & 32.67 & 1.71\\ 
\textbf{Qwen3-32B }     & 68.64 & 58.18 & 62.98  & 73.84 & 43.25 & $\downarrow$ 41.43\% & 58.14 & 2.63\\ 
\textbf{Llama3.1-8B}    & 70.99 & 62.35 & 66.39  & 74.10 & 51.15 & $\downarrow$ 30.97\% &30.61 & \textbf{0.31}\\ 
\textbf{Llama3.1-70B}   & 75.67 & 73.81 & 74.73  & 91.00 & 57.40 & $\downarrow$ 36.92\% & 54.52 & 5.43\\ 
\textbf{DeepSeek-V3.2}  & 85.23 & 85.37 & 85.30  & \textbf{96.37} & 74.91 & $\downarrow$ 22.27\% & 31.07 & 3.10\\ 
\textbf{GPT-5-mini}     & 85.81 & 83.17 & 84.47  & 92.93 & 73.88 & $\downarrow$ 20.50\% & 68.05 & 4.65\\ 
\textbf{GPT-5 }         & 88.53 & 83.59 & 85.99  & 92.48 & 75.12 & $\downarrow$ 18.77\% & 84.32 & 23.92\\ 
\textbf{Gemini-3-flash} & \textbf{90.46} & \textbf{88.24} & \textbf{89.34}  & 95.93 & \textbf{80.90} & $\downarrow$ \textbf{15.67}\% & \textbf{10.68} & 8.18\\
\bottomrule
\end{tabular}
\vspace{-4mm}
\end{table}

\begin{table*}[t]
\centering
\caption{Performance of fine-grained sub-capabilities (SCSSR). Cell colors represent performance levels (Blue: High; Red: Low).}
\label{tab:sccsr}
\vspace{-3mm}
% --- 调整设置 ---
\renewcommand{\arraystretch}{1.2} 
\setlength{\tabcolsep}{3pt}       

\resizebox{\textwidth}{!}{%

%|c|c|H|H|H|H|H|H|H|H| (后8列改为H，自动应用热力图)
\begin{tabular}{|c|c|H|H|H|H|H|H|H|H|}
\hline
\textbf{Capabilities} & \textbf{Sub-Capabilities} & 
\multicolumn{1}{c|}{\textbf{Qwen3-4B}} & 
\multicolumn{1}{c|}{\textbf{Qwen3-32B}} & 
\multicolumn{1}{c|}{\textbf{Llama3.1-8B}} & 
\multicolumn{1}{c|}{\textbf{Llama3.1-70B}} & 
% \multicolumn{1}{c|}{\textbf{GPT-4o-mini}} & 
\multicolumn{1}{c|}{\textbf{Deepseek-V3.2}} &  
\multicolumn{1}{c|}{\textbf{GPT-5-mini}} & 
\multicolumn{1}{c|}{\textbf{GPT-5}} & 
\multicolumn{1}{c|}{\textbf{Gemini-3-flash}} \\ \hline

% --- Group 1: Transformative Alignment ---
\multirow{3}{*}{\textbf{\begin{tabular}[c]{@{}c@{}}Transformative\\ Alignment\end{tabular}}} 
 & \textbf{Format Transformation} & 11.65 & 55.44 & 61.01 & 70.38 & 87.34 & 90.38 & 87.34 & 90.63 \\ \cline{2-10} 
 & \textbf{Unit Transformation}   & 11.61 & 54.19 & 52.26 & 69.03 & 85.16 & 92.26 & 90.32 & 89.68 \\ \cline{2-10} 
 & \textbf{Semantic Mapping}      & 13.71 & 51.71 & 52.96 & 66.36 & 86.60 & 80.37 & 84.11 & 86.60 \\ \hline

% --- Group 2: Reasoning & Inference ---
\multirow{4}{*}{\textbf{\begin{tabular}[c]{@{}c@{}}Reasoning \&\\ Inference\end{tabular}}} 
 & \textbf{Arithmetic Reasoning} & 9.81 & 39.51 & 39.24 & 52.59 & 67.85 & 75.48 & 77.38  & 89.37 \\ \cline{2-10} 
 & \textbf{Logical Reasoning}    & 17.39 & 63.77 & 63.77 & 72.46 & 88.41 & 85.51 & 84.06 & 92.75 \\ \cline{2-10} 
 & \textbf{Temporal Reasoning}   & 15.26 & 47.37 & 50.00 & 59.47 & 77.89 & 81.05 & 80.53 & 89.47 \\ \cline{2-10} 
 & \textbf{Multi-hop Reasoning}  & 7.73 & 29.12 & 29.64 & 34.54 & 39.69 & 52.58 & 56.44 & 78.09 \\ \hline

% --- Group 3: Distractor Robustness ---
\multirow{2}{*}{\textbf{\begin{tabular}[c]{@{}c@{}}Distractor\\ Robustness\end{tabular}}} 
 & \textbf{Attribute Distraction} & 19.49 & 60.40 & 70.26 & 75.91 & 93.96 & 85.40 & 85.77 & 91.24 \\ \cline{2-10} 
 & \textbf{Value Distraction}     & 22.26 & 60.42 & 73.26 & 79.46 & 91.42 & 87.46  & 88.37  & 94.56 \\ \hline

% --- Group 4: Evidence Faithfulness ---
\textbf{Evidence Faithfulness} 
 & \textbf{Missing Value Faithfulness} & 0.53 & 11.03 & 16.90 & 37.37 & 65.66 & 63.88 & 63.52  & 61.21 \\ \hline

% --- Group 5: Conflict Resolution ---
\multirow{3}{*}{\textbf{\begin{tabular}[c]{@{}c@{}}Conflict\\ Resolution\end{tabular}}} 
 & \textbf{Rule-based Resolution}       & 53.02 & 90.60 & 89.26 & 95.30 & 100.00 & 100.00 & 100.00 & 100.00 \\ \cline{2-10} 
 & \textbf{Constraint-based Resolution} & 54.05 & 33.33 & 64.86 & 10.81 & 70.27 & 88.29 & 81.08 & 73.87 \\ \cline{2-10} 
 & \textbf{Source-aware Resolution}     & 44.67 & 21.15 & 41.79 & 32.83 & 50.08 & 38.92 & 45.69 & 51.78 \\ \hline
\end{tabular}%
}
\vspace{-2mm}
\end{table*}

\begin{figure}
	\centering
    \includegraphics[width=0.9\linewidth]{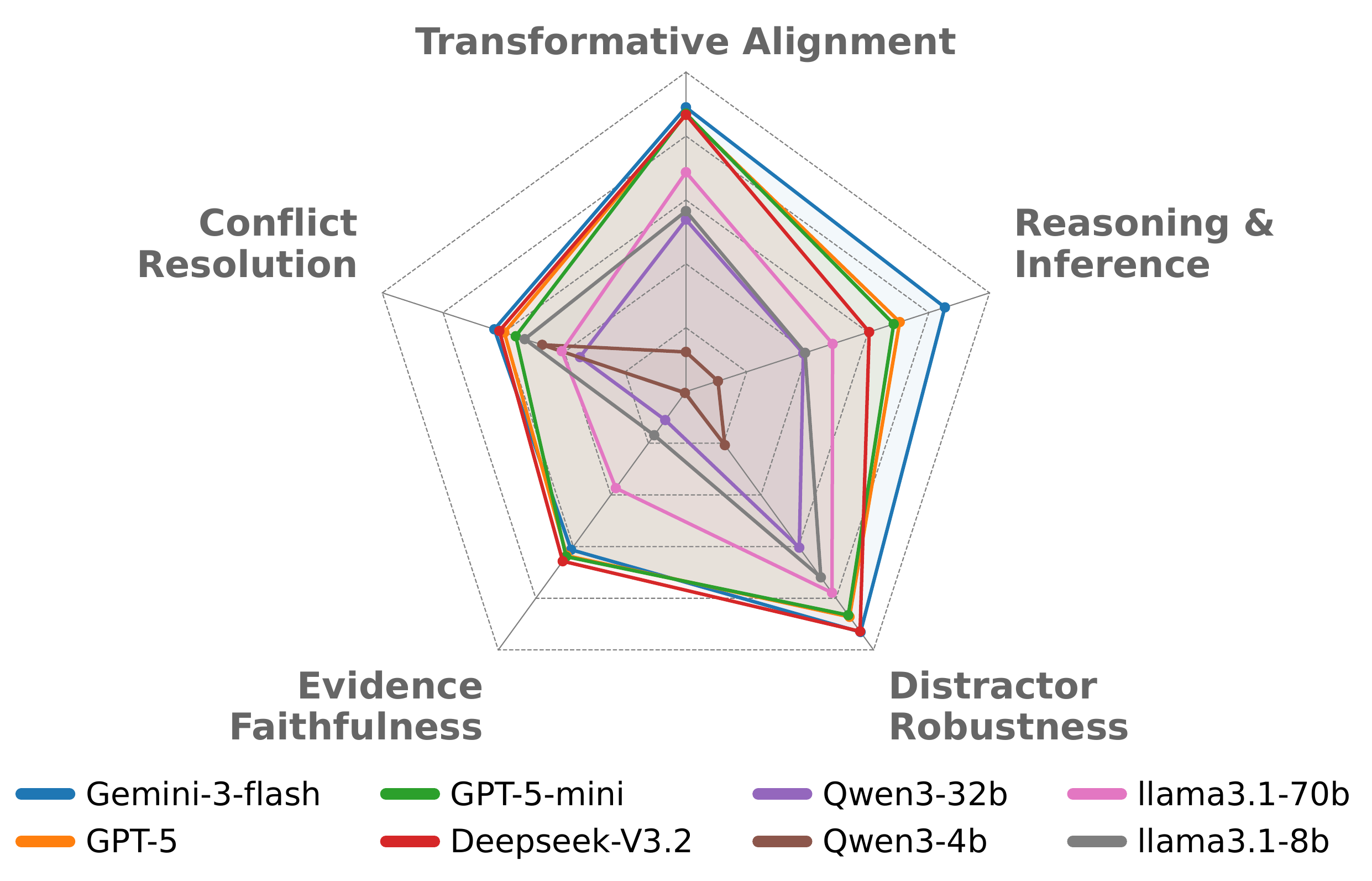}
    \vspace{-3mm}
    \caption{Performance of LLMs across five capabilities (CSSR).}
    \label{fig:radar}
    \vspace{-3mm}
\end{figure}

\noindent\textbf{Performance.} 
Table~\ref{tab:overall_performance} reports the overall performance of various LLMs on \textsf{DTBench}.
The first observation is that although leading models achieve relatively high F1 scores, the decomposed metrics $R_{\text{dir}}$ and $R_{\text{ind}}$ indicate a substantial gap between direct and indirect extraction. For example, even the Gemini-3-flash exhibits a 15.67\% drop in indirect extraction compared to direct extraction. 
% Notably, unlike open-ended text generation, Doc2Table extraction requires near-perfect accuracy to ensure the reliability of downstream analytics. These results demonstrate that current LLMs still fall short in indirect extraction, leaving considerable room for improvement.
The second observation is that scaling models within the same family (Qwen3 and Llama3.1) does not noticeably narrow the direct-indirect extraction gap. This indicates that scaling mainly improves direct extraction, while offering limited gains for capability-demanding indirect extraction, highlighting the difficulty of indirect extraction and the need for capability-aware evaluation.
 
\Takeaway 1 The persistent deficiency in indirect extraction remains a critical bottleneck for Doc2Table, as reliable downstream analytics require near-perfect Doc2Table accuracy.

In terms of time and cost, there is a clear trade-off between model capability and resource consumption. Leading models like GPT-5 and Gemini-3-flash incur high costs. In contrast, lightweight open-weight models (e.g., Llama-3.1-8B and Qwen3-4B) are highly cost-effective. Regarding efficiency, GPT-5 exhibits the highest latency, while Gemini-3-flash is efficient. Therefore, Gemini-3-flash is a good choice for scenarios demanding high efficiency and effectiveness.

% We observe a clear scaling trend: lightweight LLMs (e.g., Qwen3-4B) achieve  only 33.65\% F1, while flagship LLMs (e.g., Gemini-3-flash and GPT-5) achieve 85\%--90\% F1.
% Notably, unlike open-ended text generation, Doc2Table extraction requires near-perfect F1 to ensure the reliability of downstream analytics. Therefore, current LLMs still have considerable room for improvement on this task to meet the accuracy demands of practical use. Another observation is that, across all LLMs, performance on indirect extraction ($R_{\text{ind}}$) is consistently lower than on direct extraction ($R_{\text{dir}}$), highlighting the inherent difficulty of indirect extraction and underscoring the necessity of our capability-aware dataset construction and evaluation.
% although larger LLMs mitigate this gap more effectively. For instance, Llama 3.1-8B exhibits a 31\% drop from $R_{\text{dir}}$ to $R_{\text{ind}}$, while Gemini-3-flash shows a smaller decline of 15\%.
% \vspace{-2mm}
% \begin{findingbox}
%     Finding 1: Doc2Table extraction strongly benefits from larger LLMs, yet even SOTA LLMs leave room for improvement, with indirect extraction remaining particularly challenging.
% \end{findingbox}

\subsection{Experiment for Q3 }

% \noindent\textbf{Metrics.} 
% For a capability $c_k$, we consider only aligned cell pairs whose ground-truth cells are annotated with $c_k$. Let $TP_i^{(c_k)}$ denote the set of correctly matched cells in the $i$-th test case for capability $c_k$, and let $A_{T_i^*}^{(c_k)}$ denote the set of aligned cells in $T_i^*$ annotated with $c_k$. The Capability-Specific Accuracy (CSA) for $c_k$ is defined as follows, and Sub-capability-Specific Accuracy (SCSA) for sub-capability $c_{k,l}$ is computed analogously.
% \[
% \mathrm{CSA}(c_k)
% =
% \frac{\sum_{i=1}^N \left| TP_i^{(c_k)} \right|}
% {\sum_{i=1}^N \left| A_{T_i^*}^{(c_k)} \right|}.
% \]
\noindent\textbf{Metrics.}
To evaluate models across different Doc2Table capabilities, we define the  Capability-Specific Success Rate (CSSR) for a capability $c_k$, and the Sub-capability-Specific Success Rate (SCSSR) for a sub-capability $c_{k,l}$.
Let $C_{T_i^*}^{x}$ denote the set of all cells in $T^*$
annotated with a specific label $x$ (where $x$ can be $c_k$ or $c_{k,l}$), and let $TP_i^{x}$
denote the subset of correctly matched pairs whose ground-truth cell belongs to $C_{T_i^*}^{x}$.
The metrics are defined as: 
% $\mathrm{CSSR}(c_k)
% =
%  \sum_{i=1}^N \left| TP_i^{c_k} \right|/\sum_{i=1}^N \left| C_{T_i^*}^{c_k} \right| $, and $\mathrm{SCSSR}(c_{k,l})
% =
%  \sum_{i=1}^N \left| TP_i^{c_{k,l}} \right|/\sum_{i=1}^N \left| C_{T_i^*}^{c_{k,l}} \right| $.
\[
\mathrm{CSSR}(c_k)
=
\frac{\sum_{i=1}^N \left| TP_i^{c_k} \right|}
{\sum_{i=1}^N \left| C_{T_i^*}^{c_k} \right|}, \quad
\mathrm{SCSSR}(c_{k,l})
=
\frac{\sum_{i=1}^N \left| TP_i^{c_{k,l}} \right|}
{\sum_{i=1}^N \left| C_{T_i^*}^{c_{k,l}} \right|}.
\]

\noindent\textbf{Performance.} Figure~\ref{fig:radar} and Table~\ref{tab:sccsr} report CSSR and SCSSR results. We have the following key observations:

% \emp{(i) TA and DR are limited by fine-grained semantics.} 
\emp{(i) TA and DR.} 
Large models perform strongly on Transformative Alignment (TA) and Distractor Robustness (DR), yet fine-grained semantic challenges remain: \textit{semantic mapping} lags behind format and unit transformations, and robustness to \textit{attribute-level} distraction is lower than that to \textit{value-level} distraction. 

\Takeaway 2 Data transformation, a long-standing research topic in the database community~\cite{Auto-Transform,TDE,LiYLFT25}, has largely been internalized by modern LLMs, while future progress will depend on improved fine-grained semantic understanding.
    
% \emp{(ii) Multi-hop reasoning remains a bottleneck.}
\vspace{0.5mm}\emp{(ii) RI.}
Reasoning \& Inference capability varies substantially across models, with \textit{multi-hop reasoning} being the dominant limitation. Specifically, SCSSR analysis shows that even leading models (e.g., GPT5, Gemini-3-flash) underperform by 13\%--38\% on multi-hop reasoning relative to single-step arithmetic or logical reasoning.

\Takeaway 3 Multi-hop reasoning remains a key bottleneck; future work should focus on improving intermediate evidence retrieval, aggregation, and reasoning modularization.
    
% \emp{(iii) Faithfulness as a persistent challenge.} 
\vspace{0.5mm}\emp{(iii) EF.} 
All LLMs exhibit low CSSR of Evidence Faithfulness, even leading models like GPT-5 and Gemini-3-flash achieve only around 60\%. This indicates that LLMs have a strong tendency to fully populate the table under the given schema, rather than faithfully output \texttt{NULL} for unsupported cells. This poses substantial risks to the reliability of downstream data analysis. 

\Takeaway 4 Faithfulness remains a persistent challenge, necessitating both model-side improvement to reduce hallucination and data-side efforts to incorporate systematic, data-driven verification.

% \emp{(iv) Implicit CR remains challenging.}
\vspace{0.5mm}\emp{(iv) CR.}
As shown in Figure~\ref{fig:radar}, conflict resolution is consistently one of the weakest dimensions across models, with substantial variation across its sub-capabilities. As shown in Table~\ref{tab:sccsr}, the leading models achieve near-perfect performance on \textit{rule-based resolution} (up to 100\%), while performance drops sharply on \textit{constraint-based} and \textit{source-aware} resolution (e.g., 45.69\% for GPT-5). This indicates that applying explicit deconfliction rules is not the main challenge; rather, current LLMs struggle with implicit conflict resolution that requires reasoning about source reliability, value consistency, and implicit constraints. 

\vspace{1mm}
\Takeaway 5 LLMs struggle with implicit conflict resolution in Doc2Table when explicit rules or guidance are absent.
Future research can leverage rule mining techniques from the database and data mining communities to model explicit constraints, tightly integrating them into LLM decoding and verification.

% \subsection{Discussion: Extraction vs. Memory} 

%% file: tex/5.RelatedWork.tex
\section{Related Work}
\parhead{Structured Data Extraction} aims to extract structured information from unstructured data. Traditional information extraction (IE) decompose this process into predefined sub-tasks, such as named entity recognition and relation extraction, evolving from rule-based methods~\cite{LeeWWH13,AppeltHBIT93} to deep learning and pretrained language models-based~\cite{WuHW08,ZhengMD018,LiSZ22}. However, such pipelines are cumbersome to customize and lack flexibility for schema-driven table extraction, where target attributes may fall outside existing ontologies and required output formats may not align with predefined IE tasks~\cite{InstructIE}. 

Recent work explores the use of LLMs for structured table extraction. Evaporate~\cite{EVAPORATE} leverages LLMs and LLM-generated code to extract data from well-templated documents. Lotus~\cite{LOTUS}, PZ~\cite{PZ}, and LiteCost~\cite{Litecost} prompt LLMs to extract specific attributes, while Doctopus~\cite{Doctopus} combines LLMs with traditional IE tools to reduce cost. However, these studies do not deeply explore how effective and reliable different LLMs are on structured data extraction, especially in schema-guided Doc2Table extraction.
% Traditional information extraction (IE) decomposes extraction into predefined sub-tasks, such as named entity recognition, relation extraction, event extraction, each tied to a fixed ontology of entity and relation types. 
% However, Doc2Table requires extracting arbitrary attributes defined a schema, which often fall outside these rigid ontology. As a result, traditional IE pipelines become cumbersome to customize and insufficiently flexible for the open-ended nature of Doc2Table extraction.
% The emergence of large language models (LLMs) offers an opportunity to overcome these limitations by supporting flexible, schema-driven extraction without reliance on fixed ontologies.
% Given a document and a user instruction about extraction intent (e.g., the schema of an intended output table) as inputs, existing unstructured data analysis (UDA) systems~\cite{Doctopus,DocETL,ZenDB} leverage powerful LLMs to extract the required information and generate structured tables that follow the provided instructions.

\down\parhead {Existing Benchmark Datasets} for Doc2Table extraction largely reuse datasets originally designed for table-to-text generation, such as Rotowire~\cite{Rotowire} and Wiki40B~\cite{Wiki40B}. These datasets contain short text snippets and support relatively trivial extraction, often reducible to direct text replication.
LiveSum~\cite{LiveSum} is designed to generate summary tables from real-time competition commentary and primarily evaluates reasoning ability, but its domain and capability coverage are limited. InstructIE~\cite{InstructIE} and StructText~\cite{StructText} are synthetic benchmark datasets; however, they suffer from simplified schemata, trivial text–tuple alignments, limited validation, and do not comprehensively cover Doc2Table capabilities. More recent benchmarks, such as SemBench~\cite{SemBench} and UDA-Bench~\cite{UDABench}, target unstructured document analysis, yet still lack fine-grained and systematic evaluation of diverse Doc2Table capabilities.

%% file: tex/6.Conclusion.tex
\section{Conclusion}
\label{sec:conclusion}
In this paper, we introduce a two-level capability taxonomy for Document-to-Table (Doc2Table) extraction and present \textsf{DTBench}, the first capability-aware benchmark comprising 120 test cases and 8,811 cell-level instances across 5 major categories and 13 fine-grained subcategories. Extensive evaluations across eight mainstream LLMs demonstrate that substantial gaps persist in indirect extraction. In particular, multi-hop reasoning, evidence faithfulness, and implicit conflict resolution emerge as key bottlenecks. This study aims to foster future research toward achieving accurate Doc2Table extraction for reliable data analytics.

%% file: tex/7.Acknowledgments.tex
\section{Acknowledgments}
This work was supported in part by the NSFC under Grants No. (U23A20296 and 62025206), Zhejiang Province's ``Lingyan'' R\&D Project under Grant No. 2024C01259, and Guangdong provincial project No. 2023CX10X008. Yunjun Gao is the corresponding author  of the work.

%% file: tex/Appendix.tex
% \clearpage
\appendix

\begin{table*}[htbp]
\centering
\caption{Evaluation rubrics for generated documents (1=worst, 5=best)}
\vspace{-3mm}
\label{tab:evaluation_rubrics}
\begin{tabular}{clll} % 将 p{4.5cm} 改为 l，因为内部嵌套了自动对齐的 tabular
\toprule
\textbf{Score} & \textbf{Lexical Richness} & \textbf{Logical Consistency} & \textbf{Textual Coherence} \\ \midrule
1 & \begin{tabular}[c]{@{}l@{}}\textbf{Repetitive:}\\Minimal variety, robotic repetition\end{tabular} & \begin{tabular}[c]{@{}l@{}}\textbf{Fragmented:}\\No connections, random claims\end{tabular} & \begin{tabular}[c]{@{}l@{}}\textbf{Incoherent:}\\Difficult to follow, random jumps\end{tabular} \\ \addlinespace
2 & \begin{tabular}[c]{@{}l@{}}\textbf{Limited:} \\Simple vocabulary, narrow range\end{tabular} & \begin{tabular}[c]{@{}l@{}}\textbf{Weak:} \\Loose transitions, vague reasoning\end{tabular} & \begin{tabular}[c]{@{}l@{}}\textbf{Poor flow:} \\Jarring transitions, disconnected\end{tabular} \\ \addlinespace
3 & \begin{tabular}[c]{@{}l@{}}\textbf{Acceptable:}\\ Adequate variety, standard usage\end{tabular} & \begin{tabular}[c]{@{}l@{}}\textbf{Acceptable:}\\ Clear order, basic signaling\end{tabular} & \begin{tabular}[c]{@{}l@{}}\textbf{Acceptable:}\\ Some awkward transitions\end{tabular} \\ \addlinespace
4 & \begin{tabular}[c]{@{}l@{}}\textbf{Versatile:}\\Natural synonyms, precise terminology\end{tabular} & \begin{tabular}[c]{@{}l@{}}\textbf{Compelling:}\\Strong arguments, coherent progression\end{tabular} & \begin{tabular}[c]{@{}l@{}}\textbf{Smooth:}\\ Clear progression, minor issues\end{tabular} \\ \addlinespace
5 & \begin{tabular}[c]{@{}l@{}}\textbf{Sophisticated:} \\Rich nuances, professional mastery\end{tabular} & \begin{tabular}[c]{@{}l@{}}\textbf{Rigorous:} \\Flawless chain, seamless attribution\end{tabular} & \begin{tabular}[c]{@{}l@{}}\textbf{Seamless:} \\Natural flow, effortless transitions\end{tabular} \\ \bottomrule
\end{tabular}
\end{table*}

\begin{table}[htbp]
  \centering
  \renewcommand{\theadfont}{\small\bfseries} 
  \caption{Evaluation scores of our synthetic documents.}
  \vspace{-3mm}
  \label{tab:evaluation}
  \begin{tabular}{lcccc}
    \toprule
    \thead{Judger} & 
    \thead{Lexical \\ Richness} & 
    \thead{Logical \\ Consistency} & 
    \thead{Textual \\ Coherence} & 
    \thead{Average} \\
    \midrule
    LLM & 4.96 & 4.33 & 4.11 & 4.47 \\
    Human & 4.35 & 4.01 & 3.78 & 4.05 \\
    \bottomrule
  \end{tabular}
  \vspace{5.5mm}
\end{table}

% \begin{figure*}[h]
% % \vspace{-3mm}
% \centering
% \begin{minipage}{\linewidth}
%     \centering
%        \includegraphics[width=0.5\textwidth]{tex/radar_legend.pdf}\\
%        \vspace{-2mm}
%   \end{minipage}
% \subfigure[CSSR on \textsf{DTBench}]{
% \begin{minipage}[t]{0.48\linewidth}
% \centering
% \includegraphics[width=1\linewidth]{tex/radar_chart_DT.pdf}
%     \vspace{-3mm}
% \end{minipage}}
% \subfigure[CSSR on CounterDT]{
% \begin{minipage}[t]{0.48\linewidth}
% \centering
% \includegraphics[width=1\linewidth]{tex/radar_chart_counter.pdf}
%     \vspace{-3mm}
% \end{minipage}
% }
% \vspace{-3mm}
% \caption{\blue{Comparison of CSSR on \textsf{DTBench} and CounterDT.}}  
% \label{fig:radar_comp}
% \vspace{-2mm}
% \end{figure*}

\section{Table Collection and Processing}
\label{app:data_collect}
We collect 120 tables, covering various domains including finance, politics, economy, sports, etc. 
These tables are carefully gathered by two post-graduate students from Kaggle~\cite{kaggle}, Wikipedia~\cite{wikipedia}, and public datasets for data fusion~\cite{fusiondataset}.

To support reliable synthesis and evaluation, we require each table to include an entity name column (denoted as $a_1$) as the primary key. This aligns with real-world extraction scenarios, where information is typically organized around a natural anchor such as an entity or a temporal axis. Moreover, 
the presence of an entity name column enables unambiguous structured evidence generation and evaluation by allowing table tuples to be aligned via primary keys (see Appendix~\ref{app:alignment}).
For each table, attribute names are obtained directly from the raw tables, while column data types and constraints are manually derived or annotated when missing. We further augment the schema $S$ with attribute descriptions, data types, constraints, and representative instance examples, yielding a structured and expressive schema specification.

We additionally annotate cross-column constraints implicitly expressed by the table (e.g., $\text{Discharge\_Date} \geq \text{Admission\_Date}$), capturing latent attribute dependencies. These constraints are used to guide evidence generation during document synthesis but are not exposed during Doc2Table evaluation, as they are essential for assessing implicit conflict resolution capabilities.
To enable controlled evaluation of faithfulness-related capabilities, we deliberately remove a subset of cell values in the collected tables, and treat them as missing. The removed cells are carefully selected such that their values cannot be inferred from other cells within the same table, ensuring that correct behavior requires outputting \texttt{NULL} rather than speculative completion.
Note that we include data fusion datasets~\cite{fusiondataset} to inject source-aware resolution capability. We select the Stock, Flights, and Book datasets. Each dataset contains multiple conflicting facts for the same entities collected from heterogeneous data sources (e.g., nasdaq.com, yahoo finance, and google finance for stock information), along with a manually curated, conflict-free gold table. We treat the gold table as the ground-truth table $T^*$ for Table2Doc synthesis, while the source-annotated conflicting facts are used as inverse evidence set and can be directly fed into Step~3 in our workflow.

\section{Document Quality Evaluation}
\label{app:doc_eva}

The synthesized document $D$ should be natural and logically coherent, closely resembling real-world documents rather than artificial constructions. To assess synthesis quality, we adopt a three-dimensional evaluation framework that combines LLM-as-a-judge and human evaluation. As shown in Table~\ref{tab:evaluation_rubrics}, the evaluation considers three criteria: lexical richness, logical consistency, and coherence, each assessed using 5-point rubrics, following a similar strategy of previous work (e.g., LLMs4Synthesis~\cite{llms4synthesis}). We use DeepSeek-V3.2 as the LLM judge. In addition, we randomly sample 50 documents and have three graduate students independently evaluate them using the same rubrics. Table~\ref{tab:evaluation} reports evaluation scores from both LLM and human annotators, which confirm the high quality of the synthesized documents.

\section{Details of Row Alignment}
\label{app:alignment}

To enable cell-level comparison between a predicted table $T$ and the ground-truth table $T^*$, we first establish a row-wise alignment between the two tables. Since the number and ordering of tuples may differ, direct index-based alignment is not applicable.

\emp{Bipartite Matching Formulation.}
We formulate tuple alignment as a weighted bipartite graph matching problem. Specifically, we construct a bipartite graph $\mathcal{G} = (N_T, N_{T^*}, E)$, where $N_T$ and $N_{T^*}$ denote the sets of tuples in $T$ and $T^*$, respectively. Each edge $\langle t, t^* \rangle \in E$ represents a candidate alignment between tuple $t \in T$ and $t^* \in T^*$.

\emp{Matching Score.}
Each edge in the defined graph $\mathcal{G}$ is assigned a weight
$m(t, t^*) = \operatorname{sim}\bigl(f(t), f(t^*)\bigr)$,
where $f(\cdot)$ extracts key identifying attributes from a tuple (e.g., entity names or identifiers), and $\operatorname{sim}(\cdot)$ denotes a similarity function. To suppress spurious alignments, an edge is included only if $m(t, t^*) \ge \tau$, where $\tau$ is a predefined similarity threshold.

\emp{Optimal Alignment.}
The final row alignment is obtained by computing a maximum-weight bipartite matching over $\mathcal{G}$, yielding a one-to-one correspondence between matched tuples in $T$ and $T^*$. Unmatched tuples are treated as false positives or false negatives in subsequent cell-level evaluation.

\begin{table}[t]
\centering
\caption{Comparison between TableLLM (with Llama-3.1-8B as backbone) and Vanilla Llama-3.1-8B.}
\vspace{-3mm}
\label{tab:tablellm}
\begin{tabular}{l c c c c c}
\toprule
\textbf{Model} & \textbf{P} & \textbf{R} & \textbf{F1} & $\mathbf{R_{\text{dic}}}$ & $\mathbf{R_{\text{ind}}}$ \\
\midrule
TableLLM-8b   & 54.26 & 33.37 & 41.32 & 45.24 & 22.09 \\
Llama-3.1-8b  & 70.99 & 62.35 & 66.39 & 74.10 & 51.15 \\
\bottomrule
\end{tabular}
\vspace{4mm}
\end{table}

\section{Experiments on TableLLM}
\label{app:tablellm}

We conduct an additional experiment on \textsf{DTBench} using a recent TableLLM~\cite{tablellm}, which is fine-tuned on Llama-3.1-8B. Table~\ref{tab:tablellm} presents the results compared with the vanilla Llama-3.1-8B. As observed, TableLLM underperforms the vanilla Llama-3.1-8B on \textsf{DTBench}.

\section{Experiments on Model Bias}
\label{app:new_DTBench}
Since Grok-4-fast is used as the default backbone of the  document synthesizer, a natural concern is whether our experimental findings are biased toward extractor LLMs whose architectures are more aligned with the synthesizer. However, we argue that such bias does not materially affect our conclusions (see discussion in Section~\ref{subsec:bench}). Also, as shown in Table~\ref{tab:overall_performance}, LLMs across diverse architectures perform consistently with their well-recognized capabilities, showing no disproportionate advantage for any specific model family. To further demonstrate this, we resynthesize the entire dataset utilizing GPT-4o-mini as the synthesizer. We compared the performance of 5 representative LLMs on the resynthesized dataset (New F1) against the results on the original dataset generated using Grok-4-fast (Original F1), as shown in Table~\ref{tab:bias}.
We have two key observations. (i) Rank Consistency: The relative ranking of the LLMs remains absolutely consistent.
(ii) No ``Intra-Family'' Advantage: Utilizing GPT-4o-mini as the synthesizer creates a highly favorable testing condition for GPT-5-mini due to their shared linguistic patterns. However, Gemini-3-flash strictly maintains its lead, and the marginal F1 increase for GPT-5-mini (+1.36) is almost identical to that of Llama-3.1-70B (+1.30). This demonstrates that writing styles cannot enable a model to leapfrog stronger competitors from a higher tier in Doc2Table.

 \begin{table}[t]
\centering
\caption{Comparison of Original and New F1 Scores}
\vspace{-3mm}
\label{tab:bias}
\begin{tabular}{l c c c}
\toprule
\textbf{Extractor LLM} & \textbf{Original F1} & \textbf{New F1} & \textbf{Rank} \\
\midrule
\textbf{Gemini-3-flash} & 89.34 & 87.28 & 1 $\rightarrow$ 1 \\
\textbf{Deepseek}       & 85.30 & 86.03 & 2 $\rightarrow$ 2 \\
\textbf{GPT-5-mini}     & 84.47 & 85.83 & 3 $\rightarrow$ 3 \\
\textbf{Llama-3.1-70B}  & 74.73 & 76.03 & 4 $\rightarrow$ 4 \\
\textbf{Qwen3-32B}      & 62.98 & 58.80 & 5 $\rightarrow$ 5 \\
\bottomrule
\end{tabular}
\end{table}